# Perfect spin-triplet pairing in two-dimensional Ising superconductors purified by indirect excitons

Chuanyi Zhang[1], Yu Li[2], Ping Cui[3], and Zhenyu Zhang[3*]

[1] Henan Key Laboratory of Quantum Materials and Energy and School of Future Technology, Henan University, Kaifeng 475004, China;

[2] Institute of Quantum Materials and Physics, Henan Academy of Sciences, Zhengzhou 450046, China;

[3] International Center for Quantum Design of Functional Materials (ICQD), University of Science and Technology of China, Hefei 230026, China, and Hefei National Laboratory, University of Science and Technology of China, Hefei 230088, China.

[*]Corresponding author: zhangzy@ustc.edu.cn

**Abstract**

Much research effort has been devoted to interfacial or two-dimensional (2D) superconductors, but the underlying pairing mechanisms and pairing symmetries are highly controversial in most cases. Here we propose an innovative approach to probe the pairing symmetry of 2D superconductors, based on a van der Waals heterostructure consisting of a prototypical 2D Ising superconductor coupled with a 2D hole gas through an insulating spacer. We first show that, by tuning the Coulomb attraction between the superconducting and hole layers, the gap of the corresponding indirect exciton insulators is tuned as well, resulting in contrasting manifestations of the distinct superconducting channels with spin-singlet ($s$-, extended $s$-, and $d$-wave) and spin-triplet ($p$- and $f$-wave) pairings. Strikingly, we find that the application of in-plane magnetic fields can suppress all other channels while selecting the spin-triplet $p$-wave channel to be the pure superconducting state, thus providing an ideal and practical platform for realizing highly desirable topological superconductivity. Such an approach can also be readily extended to other types of superconducting systems, offering unprecedented opportunities to probe the microscopic mechanisms of unconventional superconductivity.

**Main**

As one of the macroscopic quantum phenomena, superconductivity occupies a prominent position in condensed matter physics, renowned for its magnificent manifestations and immense technological potentials. Tremendous efforts have been devoted to discovering new types of superconductors and deciphering their underlying pairing mechanisms under diverse conditions[1,2]. More recently, various fabrication techniques, such as mechanical exfoliation and molecular beam epitaxy, have been introduced into the field of superconductivity, enabling realization of highly crystalline two-dimensional (2D) superconductors[3,4]. Compelling examples include few-atomic-layer metals on semiconducting substrates[5], few-layer transition metal dichalcogenides[6,7,8,9,10], and magic-angle graphene[11,12]. Yet to date, the underlying pairing mechanisms, pairing strengths, and pairing symmetries remain open in most cases.

Separately, as a close analogue to superconductivity, an exciton insulator represents a new insulating ground state of matter that arises from spontaneous formation of bound electron-hole pairs due to Coulomb attraction. In this field, spatially separated (or indirect) excitons have attracted substantial attention owing to their long lifetime. For example, indirect exciton insulators have been observed in coupled semiconductor quantum wells[13], including demonstration of their spontaneous coherence[14]. As highly crystalline 2D materials arise, new platforms based on van der Waals (vdW) heterostructures have also become available for investigation of indirect exciton insulators[15,16,17,18,19,20,21], and in particular, their superior interfacial structures with atomic scale precision allow to explore intriguing properties such as strongly correlated exciton fluids and exotic optoelectronic transport[20,21]. It is worthwhile to note that, unlike the superconducting case where the dominant pairing mechanisms are heavily debated, here the electron-hole pairing is well defined to be of Coulomb attraction.

Here we propose a conceptually innovative approach to study superconductivity by coupling a superconductor with a probing indirect exciton insulator using a vdW heterostructure. As the first demonstration of this powerful approach, we take the 2D Ising superconductor as a prototypical example, coupled with a 2D hole gas through an insulating spacer (see Fig. 1). We show that, by tuning the Coulomb attraction between the superconducting and hole layers, the gap of the corresponding indirect exciton insulators is tuned as well, resulting in contrasting manifestations of the distinct superconducting channels with spin-singlet ($s$-, extended $s$-, and $d$-wave) and spin-triplet ($p$- and $f$-wave) pairings. Strikingly, we find that the application of in-plane magnetic fields can suppress all other channels while selecting the spin-triplet $p$-wave channel to be the long-sought pure superconducting state, thus providing an ideal and yet practical platform for realizing highly desirable topological superconductivity. This approach is also expected to be broadly applicable to other 2D or interfacial superconductors[22,23,24,25], potentially offering unprecedented insights into the superconducting pairing mechanisms and related properties.

As a constituent component of the architecture, the concept of Ising superconductivity[7] was first proposed and experimentally observed in $MoS_2$ with substantial subsequent extensions to other systems[26,27,28]. In essence, the spin-orbit coupling associated with the non-centrosymmetric monolayer transition metal dichalcogenides pins the electron spins to the out-of-plane direction, and such an Ising-type spin-orbit coupling enables the in-plane critical magnetic field to be far beyond the Pauli limit[29]. A better understanding of the prototypical 2D Ising superconductor will not only attest the significance of the present approach, but such Ising superconducting systems themselves may also harbor rich new physics, as exemplified by the recent discoveries of Fulde-Ferrell-Larkin-Ovchinnikov states[30,31].

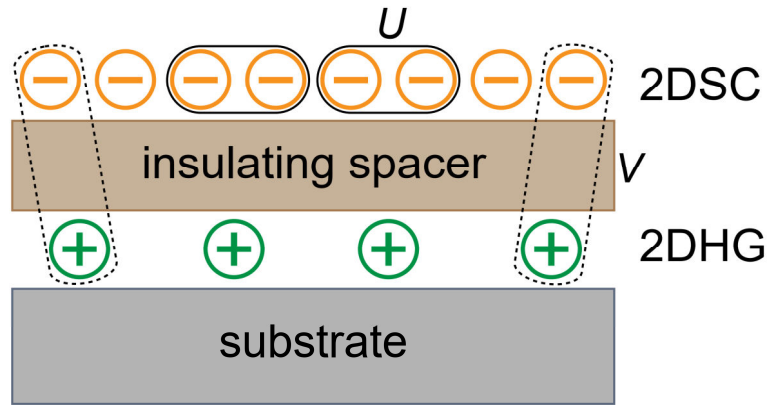

**Fig. 1: Illustration of the architecture coupling a two-dimensional superconductor (2DSC) with a two-dimensional hole gas (2DHG).** The 2DSC and 2DHG are separated by an insulating spacer, and the whole device is supported on a proper substrate. The Cooper pairs with strength $U$ and indirect excitons with strength $V$ are represented by the solid and dotted frames, respectively.

When constructing the Hamiltonian for the proposed architecture in Fig. 1, we choose monolayer MoS$_2$ and WS$_2$ as the 2D electron gas and hole gas layers, respectively, separated by a hexagonal boron nitride (h-BN) insulating spacer with tunable thickness as discussed later. The electron density in MoS$_2$ can be experimentally adjusted to the order of $10^{18}$ m$^{-2}$, which can be achieved through electrostatic induction by using an ionic liquid gate[6,7]. The hole density in WS$_2$ is tuned by the same approach of gating[32] to be substantially lower than that of electrons, taken to be about half of the electron density. The Hamiltonian has the following form in such a system,

$$\mathcal{H} = \sum_{k\tau s}\left(\varepsilon_{k\tau s}e_{k\tau s}^{\dagger}e_{k\tau s} - \xi_{k\tau}h_{k\tau}^{\dagger}h_{k\tau}\right) + \sum_{k\tau ss'}(\tau\beta\boldsymbol{e}_z - \boldsymbol{\Lambda})\cdot\boldsymbol{\sigma}_{ss'}e_{k\tau s}^{\dagger}e_{k\tau s'}$$

$$+ \frac{1}{2}\sum_{kk'q\tau ss'}U_q\,e_{k+q\tau s}^{\dagger}e_{k'-q\bar{\tau}s'}^{\dagger}e_{k'\bar{\tau}s'}e_{k\tau s}$$

$$- \sum_{kk'q\tau s}V_q\,e_{k+q\tau s}^{\dagger}h_{k'-q\bar{\tau}}h_{k'\bar{\tau}}^{\dagger}e_{k\tau s'}$$

$$(1)$$

where $e_{k\tau s}^{\dagger}$ ($h_{k\tau}^{\dagger}$) is a creation operator for an electron (a hole) with momentum $\boldsymbol{k}$ and spin $s$ near valley $\tau$ in the MoS$_2$ (WS$_2$) layer, and the corresponding energy is represented by $\varepsilon_{k\tau s}$ ($\xi_{k\tau}$) with the chemical potential included. The coefficient $\tau = \pm1$ for valley K and K′, respectively, and the index $\tau$ represents the valleys, with $\bar{\tau}$ being the opposite valleys. The coefficient $\beta$ denotes the strength of the Ising-type spin-orbit coupling in the electron layer, and $\boldsymbol{\sigma}_{ss'}$ is the vector of Pauli matrices for spin. Here we note that, because of the large spin-orbit coupling of the valence band in WS$_2$, the holes with spin up and spin down configurations in the respective K and K$'$ valleys are well separated, allowing us to ignore their spin degrees of freedom in subsequent treatment. The Zeeman field $\boldsymbol{\Lambda}$ is associated with an in-plane magnetic field. The parameter $U_q$ parametrizes the general pairing interaction between electrons in opposite valleys of K and K$'$, with both the singlet and triplet pairing channels allowed by the symmetry of the structure. The exciton insulating state will emerge due to the interlayer Coulomb attraction of strength $V_q$. By employing the contact-interaction approximation[33], $V_q$ near given valleys is independent of the momentum, i.e., $V_q \approx V$, determined by the interlayer distance. In the proposed double-layer heterostructure, electrons and holes are coupled via interlayer Coulomb attraction, resulting in the formation of indirect excitons with long lifetime. These excitons are stabilized when the Coulomb potential is sufficiently large, ensuring a binding energy greater than 0.1 eV[18,19,20]. Due to the significant binding energy, the indirect excitons in this structure remain stable against thermal dissociation at temperatures well above room temperature, over a broad range of electron and hole densities[34,35].

The superconducting pairing symmetry is closely associated with the crystal structure, and possible pairing phases can be categorized by the group theoretical approach[36]. For monolayer MoS$_2$, all the possible spin-singlet and triplet pairing channels are obtained according to the irreducible representations of the point group C$_{3v}$ [37,38]. Specifically, we consider all the possible pairing forms in such a system, including the s wave, extended s wave (es wave), p+ip or p-ip wave (p±ip wave), d+id or d-id wave (d±id wave), and f wave. By applying the mean-field approximation to the above Hamiltonian, the gap functions are obtained by minimizing the free energy of the system. To assess the validity of the mean-field treatment, we invoke the Ginzburg criterion[39,40] and find that the contributions of quantum fluctuations are insignificant in the regime studied (see Supplementary information for details).

When the time-reversal symmetry is preserved in this system, the pairing matrix in the Ising superconductor (ISC) has the general form $\Delta(\mathbf{k}) = [\Delta_s(\mathbf{k})\sigma_0 + d_z(\mathbf{k})\sigma_z]i\sigma_y$, where $\Delta_s(\mathbf{k})$ and $d_z(\mathbf{k})$ represent the spin-singlet and spin-triplet pairing order parameters, respectively. Note that the $d_{x,y}(\mathbf{k})$ components are missing due to the constraint of the large Ising spin-orbit coupling in $MoS_2$ [37,38]. For the s-wave pairing channel, the ISC gap is a constant if $V$ is weaker than the threshold for the formation of exciton insulators. As $V$ increases, more electrons in $MoS_2$ tend to combine with holes in $WS_2$ to form excitons, widening the excitonic gap, while fewer electrons are left for Cooper pairs, resulting in a decrease in the superconducting gap. Consequently, delicate competitions arise between the superconducting and exciton insulating states for the given electron and hole densities in the $MoS_2$ and $WS_2$ layers. Strikingly, with physical parameters applicable for the $MoS_2$/h-BN/$WS_2$ system, we find that, at $V \approx 4$ meV, the superconducting gap of the s-wave pairing channel abruptly vanishes, marking the occurrence of a first-order phase transition (FOPT), as shown in Fig. 2a. Simultaneously, a sudden change takes place in the gap of the exciton insulator (see the inset in Fig. 2a). It is important to note that the gaps of exciton insulators with spin-singlet and spin-triplet textures are nearly identical; accordingly, these excitons with different spin textures exert similar influences on the Cooper pairs. Similar variations in both the superconducting and excitonic gaps are also observed for the $d\pm id$-wave and $f$-wave pairing phases, even though their pairing gaps are different for a given strength $V$. Furthermore, the critical values of $V$ for the FOPT points vary significantly among the $s$-wave, $d\pm id$-wave, and $f$-wave pairing channels, as evidenced in Figs. 2a, 2c, and 2d. Notably, the smallest critical value is observed for the $s$ wave, while the largest value corresponds to the $f$ wave.

In the proposed van der Waals heterostructure $MoS_2$/h-BN/$WS_2$, electrons and holes reside in different layers. The Cooper pairs exist between electrons in the $MoS_2$ layer, while indirect excitons are formed between electrons in the $MoS_2$ and holes in the $WS_2$ layer due to Coulomb attraction. Hence electrons within the Cooper pairs and those within the indirect excitons actively compete with each other in the $MoS_2$ layer. Since the densities of electrons and holes are fixed, the competition between the electrons also transforms into interplay and competition between the Cooper pairs and indirect excitons. Specifically, the Cooper pairs dominate when the Coulomb attraction is weak. However, as the Coulomb attraction increases, more electrons are dragged to form excitons with the holes, leading to the increase in the exciton insulating gap. This, in turn, results in the destruction of more Cooper pairs and a corresponding decrease in the superconducting gap. The variations of the superconducting and exciton insulating gaps can be reflected in the total free energy of the system. Upon minimization of the free energy, FOPT can be identified as the Coulomb attraction strength increases. Such a transition is signified by a sharp jump of the superconducting gap to zero, reflecting the destruction of superconductivity as shown in Figs. 2a, 2c, and 2d.

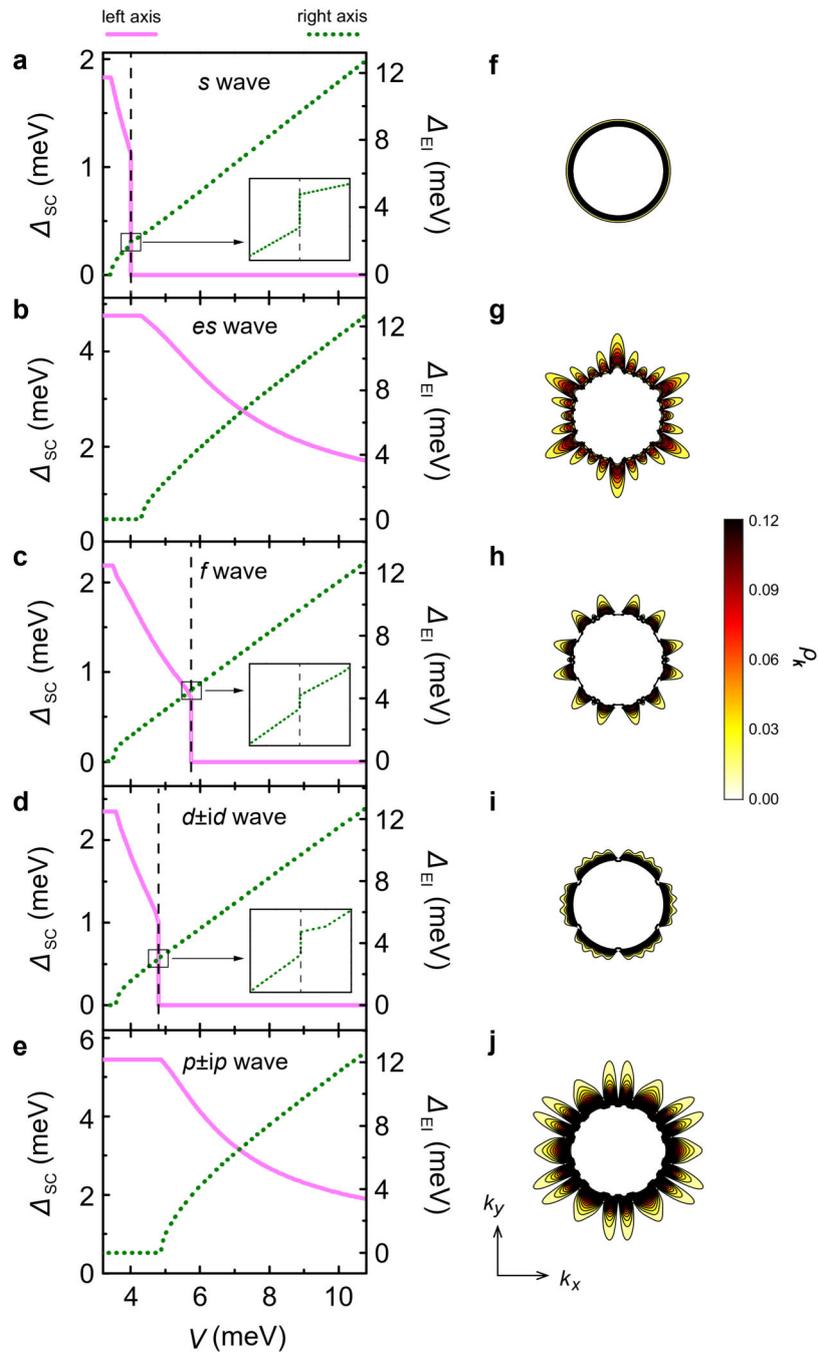

**Fig. 2: Gaps of different superconducting channels and exciton insulators and condensate densities for different Cooper pairing channels. a-e** Gaps of the superconducting channels ($\Delta_{SC}$) and exciton insulators ($\Delta_{EI}$) at increasing Coulomb attraction strength ($V$). The insets in **a**, **c**, and **d** are the zoomed-in views. **f-j** Corresponding variations in the condensate densities around the Fermi surface centered at valley K (K') for the different Cooper pairing channels. Model parameters are $V = 3.8$ meV for **f**, **h**, and **i**, and $V = 5.4$ meV for **g** and **j**.

When the *es*-wave pairing is considered, we find that the superconducting gap also decreases with increasing $V$, as shown in Fig. 2b. Nevertheless, in strong contrast with the *s* channel, the decay of this channel gradually slows down, and there is no FOPT even for relatively large $V$. This observation indicates that the superconductivity persists even when the exciton bindings become quite strong, albeit with lower superconducting transition temperatures ($T_c$). Similar behaviors are also observed for the $p\pm ip$-wave pairing in Fig. 2e.

To gain insights into the different dependences of the pairing channels on the exciton binding strength, we analyze the condensate densities of the Cooper pairs, given by $\rho_k = \sum_s |< e^{\dagger}_{kKs} e^{\dagger}_{-kK'\bar{s}} > |^2$ in reciprocal space[41,42]. As depicted in Figs. 2f-2j, electrons participating in the Cooper pairs predominantly originate from regions near the Fermi surface, and $\rho_k$ exhibits distinctly different characteristics for distinct pairing channels. Three factors influence the superconducting channels: the density distribution of the Cooper pairs in reciprocal space, *s*-wave nature of the exciton insulating gap, and electron and hole densities. Specifically, for given electron and hole densities, the Cooper pairs are more susceptible to destruction if their constituent electrons are located on or near the Fermi surface. As the Coulomb attraction increases, these electrons are more easily pulled into formation of indirect excitons due to the *s*-wave nature of the exciton insulators. On the other hand, Cooper pairs consisting of electrons further away from the Fermi surface (such as those in the *es*-wave and $p \pm ip$ wave channels) are more robust against the destruction associated with enhanced exciton formation. As a consequence, the FOPT takes place for the *s*, $d \pm id$, and *f* waves (Figs. 2f, 2h, and 2i), while the *es* and $p \pm ip$ waves survive and there are no FOPTs for the given electron and hole densities (Figs. 2g and 2j). In terms of total free energies (with the temperature fixed at 0.2 and 1.0 K, respectively), we confirm that the variations in the free energy align well with the above qualitative analyses. Specifically, the free energy for the *s*-wave pairing is the highest for a given Coulomb attraction, while the *es* and $p \pm ip$ waves have lower free energies and become nearly degenerate at larger $V$ (see details in Supplementary information). Furthermore, when we increase the hole density and keep the electron density fixed, we find that all the superconducting channels can be destroyed with increasing Coulomb attraction, especially when the electron and hole densities are comparable (such as in the cases of Refs. [19,20]).

Next, we apply a magnetic field in the *xy* plane, and examine how the different channels respond. Here, the time-reversal symmetry is broken, leading to finite in-plane components $\boldsymbol{d}_{x,y}$ for the spin-triplet order parameter. Besides, the $\boldsymbol{d}_{x,y}$ vectors are imaginary, allowing for a mixing of pairing channels that belong to different irreducible representations[43,44]. In particular, the spin-triplet $p$ channel induced by the magnetic field is characterized by pairing of electrons with parallel spins, and the free energy is minimized in the properly mixed state of $\Delta_{mix}(\boldsymbol{k}) = \left[\Delta_{es}(\boldsymbol{k})\sigma_0 + (\boldsymbol{d}(\boldsymbol{k}) \cdot \boldsymbol{\sigma})_{x,y}\right]i\sigma_y$, where $\Delta_{es}(\boldsymbol{k})$ and $d_{x,y}(\boldsymbol{k})$ represent the gap parameters of the dominant $es$ and $p$ waves, respectively. For weak magnetic fields, both the $es$ and $p$ components survive, and the gap of the $p$-wave channel is larger than that of the $es$-wave channel. Both gaps decrease smoothly when the Coulomb attraction increases, as shown in Fig. 3a. Strikingly, when the magnetic field is sufficiently large, we find that the $es$-wave channel can be fully suppressed by the Coulomb attraction as shown by the emergence of the FOPT in Fig. 3b, thereby purifying the mixed pairing phases to the perfect $p$-wave state. When the FOPT takes place, most of the electrons in the $es$-wave channel are dragged to form indirect excitons, while a small fraction of these electrons joins the $p$-wave channel, as indicated by the slight increase in the $p$-wave gap at the FOPT (Figs. 3b and 3c). Here we note that, at a higher magnetic field, the FOPT appears at a lower Coulomb interaction, but in practice the strength of the magnetic field is limited by the ISC nature.

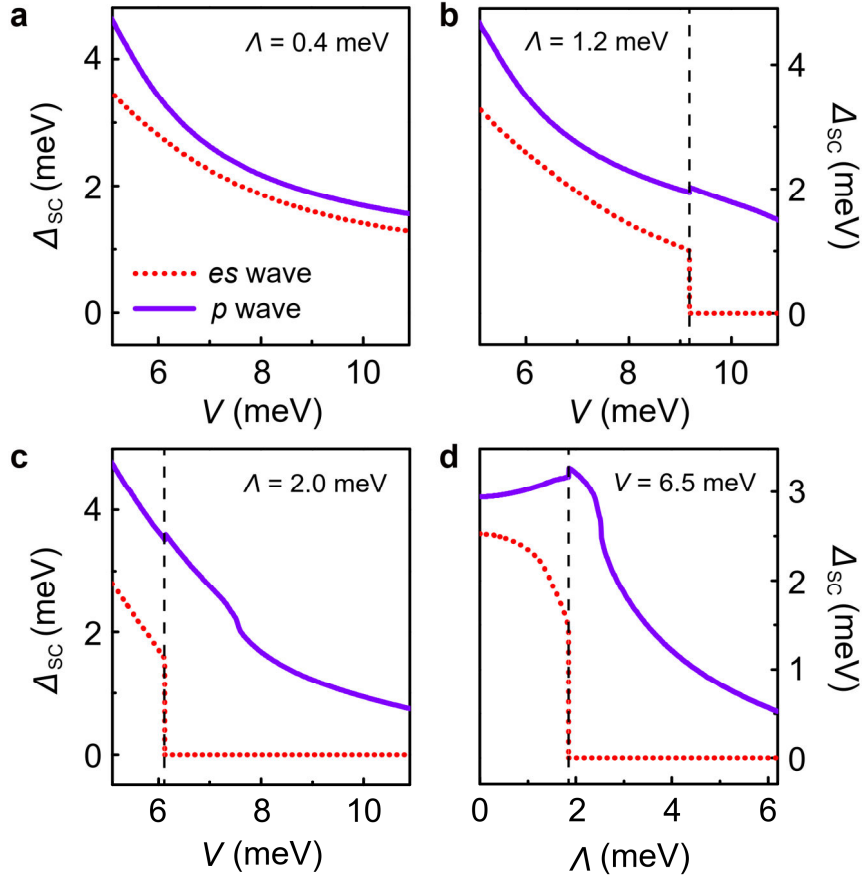

**Fig. 3: Purification of the perfect *p*-wave channel by the Coulomb attraction and in-plane magnetic field. a-c** Gaps of the *es*-wave and *p*-wave pairings ($\Delta_{SC}$) at increasing Coulomb attraction strength ($V$) with different Zeeman fields ($\Lambda$). **d** $\Delta_{SC}$ for the *es* and *p* waves at increasing $\Lambda$ with $V = 6.5$ meV.

Another very intriguing and distinctly novel aspect is revealed at a given Coulomb attraction strength but with varying magnetic field. As shown in Fig. 3d, even at relatively weak magnetic fields, the gap of the *es*-wave channel decreases rapidly, while the gap of the *p*-wave channel increases slowly, exhibiting a small but appreciable upturn discontinuity at the FOPT of the *es*-wave channel. These observations are in clear contrast with the conventional wisdom that a magnetic field reduces the superconducting gap. Qualitatively, these anomalous observations can be attributed to the picture that some electrons escaping from the *es*-wave channel are able to join the *p*-wave channel. In this regime, competition persists between the Cooper pairs and indirect excitons as the Coulomb attraction increases, even in the presence of an in-plane magnetic field. The magnetic field has a significant impact on the behaviors of the Cooper pairs, and ultimately, the combined effects of the Coulomb attraction and magnetic field give rise to the emergence of pure *p*-wave superconductivity.

**Discussion & Perspective**

Since our primary objective is to use the exciton insulating phase to differentially tune the superconducting channels, we must restrict the exciton phases to be of *s*-wave in nature. In fact, this type of exciton insulators has also been identified experimentally[18,19,20]. On the other hand, for excitons based on 2D systems, the orbital part can also adopt other types of features such as *p*-wave excitons[45,46]. As a potential future study, it would be intriguing to use similar heterostructures but with *s*-wave superconductivity to explore the *p*-wave nature of the exciton insulators.

Though the pairing mechanisms and pairing symmetries of 2D superconductors are still under active debate in most cases, our present study offers an innovative approach to probing the evolution of superconductivity and competing pairing channels under the modulation of indirect exciton insulators coupled to the 2D superconductors. Crucially, this enabling approach allows to select a pure spin-triplet channel in the presence of an in-plane magnetic field. Moreover, the proposed architecture based on van der Waals heterostructures is physically realistic and has in fact been experimentally demonstrated by using state-of-the-art fabrication techniques[19,20]. Specifically, monolayered $MoS_2$ and $WS_2$ can be produced by using physical or chemical deposition[3], and a few layers of h-BN can be utilized to serve as an insulating spacer. By adjusting the layer thicknesses of wedged h-BN (for example, using a vicinal spacer with ordered multilayers of h-BN as conceptually presented in Ref. [20]), we can effectively tune the strength of the Coulomb attraction $V$. In fact, such a sample, once fabricated, is itself also a novel superconducting device[47], because intriguing supercurrents may be sustained at the boundaries of neighboring terraces. Furthermore, the electron and hole densities in such a device can be controllably and separately tuned by double-side ionic gating[32], giving rise to the coexistence of and competition between 2D superconductors and indirect exciton insulators. The present realization scheme of pure $p$-wave superconductivity without involving ferromagnetic components[48,49,50] will provide a unique platform for investigating potential topological properties[50,51,52]. Such spin-triplet supercurrents can be exploited for developments of novel superconducting spintronic devices. From a broader perspective, our proposed heterostructure for modulating 2D Ising superconductivity with indirect excitons can be extended to other types of superconducting systems, offering new angles to probe the microscopic mechanisms of superconductivity, including that of strongly correlated electron systems.

## Acknowledgments

This work is supported by the Innovation Program for Quantum Science and Technology (Grant No. 2021ZD0302800), the National Natural Science Foundation of China (Grant Nos.12374458 and 11974323), the Strategic Priority Research Program of Chinese Academy of Sciences (Grant No. XDB0510200), and the Anhui Provincial Key Research and Development Project (Grant No. 2023z04020008).

heterostructures. *Science* **376**, 406-410 (2022).

# Supplementary information

## Perfect spin-triplet pairing in two-dimensional Ising superconductors purified by indirect excitons

Chuanyi Zhang[1], Yu Li[2], Ping Cui[3], and Zhenyu Zhang[3]*

[1] Henan Key Laboratory of Quantum Materials and Energy and School of Future Technology, Henan University, Kaifeng 475004, China;

[2] Institute of Quantum Materials and Physics, Henan Academy of Sciences, Zhengzhou 450046, China;

[3] International Center for Quantum Design of Functional Materials (ICQD), University of Science and Technology of China, Hefei 230026, China, and Hefei National Laboratory, University of Science and Technology of China, Hefei 230088, China.

*Corresponding author: zhangzy@ustc.edu.cn

## 1. Model

According to the crystal symmetry of monolayer $MoS_2$, the spin-singlet and spin-triplet superconducting gap functions can be classified by the irreducible representations of the point group $C_{3v}$, and possible paring phases are expressed by a set of the basis functions (as presented in Refs. [S1,S2]). The interaction terms in the Hamiltonian given in the main text are treated with the mean-field approximation based on possible gap functions of the superconductor and exciton insulator.

The interaction between electrons has the form,

$$\frac{1}{2}\sum_{kk'q\tau ss'} U_q\, e^{\dagger}_{k+q\tau s} e^{\dagger}_{k'-q\bar{\tau}s'} e_{k'\bar{\tau}s'} e_{k\tau s},$$

(S1.1)

which can be approximatively expressed as

$$\frac{1}{2}\sum_{kk'\tau ss'} U_{ss'}(\boldsymbol{k},\boldsymbol{k}') <e^{\dagger}_{k\tau s} e^{\dagger}_{-k\bar{\tau}s'}> e_{k'\bar{\tau}s'} e_{-k'\tau s} + \frac{1}{2}\sum_{kk'\tau ss'} U_{ss'}(\boldsymbol{k},\boldsymbol{k}')\, e^{\dagger}_{k\tau s} e^{\dagger}_{-k\bar{\tau}s'} <e_{k'\bar{\tau}s'} e_{-k'\tau s}$$
$$> -\frac{1}{2}\sum_{kk'\tau ss'} U_{ss'}(\boldsymbol{k},\boldsymbol{k}') <e^{\dagger}_{k\tau s} e^{\dagger}_{-k\bar{\tau}s'}> <e_{k'\bar{\tau}s'} e_{-k'\tau s}>.$$

(S1.2)

In Eq. (S1.2), the spin-singlet and spin-triplet gap functions have the form

$$\Delta_{ss'}(\boldsymbol{k}) = \sum_{\boldsymbol{k}'\tau} U_{ss'}(\boldsymbol{k},\boldsymbol{k}') < e_{\boldsymbol{k}'\bar{\tau}s'}e_{-\boldsymbol{k}'\tau s} > .$$

$$(S1.3)$$

For the indirect exciton insulator, with the same method applied, we obtain

$$\sum_{\boldsymbol{k}\boldsymbol{k}'q\tau s} V_q\, e^\dagger_{\boldsymbol{k}+q\tau s} h_{\boldsymbol{k}'-q\bar{\tau}} h^\dagger_{\boldsymbol{k}'\bar{\tau}} e_{\boldsymbol{k}\tau s}$$

$$\approx V \sum_{\boldsymbol{k}\boldsymbol{k}'\tau s} < e^\dagger_{\boldsymbol{k}\tau s} h_{-\boldsymbol{k}\bar{\tau}} > h^\dagger_{\boldsymbol{k}'\bar{\tau}} e_{-\boldsymbol{k}'\tau s} + V \sum_{\boldsymbol{k}\boldsymbol{k}'\tau s} e^\dagger_{\boldsymbol{k}\tau s} h_{-\boldsymbol{k}\bar{\tau}} < h^\dagger_{\boldsymbol{k}'\bar{\tau}} e_{-\boldsymbol{k}'\tau s}$$

$$> -V \sum_{\boldsymbol{k}\boldsymbol{k}'\tau s} < e^\dagger_{\boldsymbol{k}\tau s} h_{-\boldsymbol{k}\bar{\tau}} >< h^\dagger_{\boldsymbol{k}'\bar{\tau}} e_{-\boldsymbol{k}'\tau s} > ,$$

$$(S1.4)$$

with the gap function of the exciton insulator,

$$\Delta_{EI} = V \sum_{\boldsymbol{k}\tau s} < e^\dagger_{\boldsymbol{k}\tau s} h_{-\boldsymbol{k}\bar{\tau}} > .$$

$$(S1.5)$$

It should be noted that the contact-interaction approximation is applied to deal with the statically screened Coulomb attraction, and the effective interaction parameter is almost independent of the momentum [S3]. Hence, only the *s*-wave pairing phase appears in the exciton insulator.

With the mean-field approximation, the Hamiltonian of the system is simplified and can describe all the possible superconducting pairing phases. The total free energy can be expressed as,

$$\mathcal{F} = \frac{1}{2} \sum_{kk's_1's_2's_1's_2'} \Delta^*_{s_1's_2'}(\boldsymbol{k}') U^{-1}_{s_1s_2s_1's_2'}(\boldsymbol{k},\boldsymbol{k}') \Delta_{s_1s_2}(\boldsymbol{k}) + \sum_{\tau s} \frac{|\Delta_{EI\tau s}|^2}{V} + 2 \sum_{\boldsymbol{k}\tau s} \varepsilon_{\boldsymbol{k}\tau s}$$

$$-\frac{1}{\gamma} \sum_{\boldsymbol{k}} \sum_{j=1}^{10} \ln\left(1 + e^{-\gamma E_j(\boldsymbol{k})}\right),$$

(S1.6)

where $\gamma$ is the reciprocal temperature, and $E_j(\boldsymbol{k})$ is the $j^{th}$ eigenvalue of the pairing Hamiltonian. Using Eq. (S1.6), we can obtain the gaps of the superconductor and exciton insulator.

## 2. Ginzburg criterion

The mean-field method is employed to address the gaps of superconducting and exciton insulating phases in our proposed van der Waals heterostructure. In this context, it is crucial to examine the validity of the mean-field approximation in the vicinity of the superconducting transition temperature $T_c$. The Ginzburg-Landau free energy is expressed as,

$$\text{F}(\Delta_{SC}, \Delta_{EI}) = \text{F}_{SC}(\Delta_{SC}) + \text{F}_{EI}(\Delta_{EI}) + \text{F}_{SE}(\Delta_{SC}, \Delta_{EI}),$$

(S2.1)

where

$$\text{F}_{SC}(\Delta_{SC}) = \int d^d x (a_{SC}|\Delta_{SC}|^2 + \frac{1}{2} b_{SC}|\Delta_{SC}|^4 + \text{K}_{SC}|\nabla\Delta_{SC}|^2),$$

(S2.2)

$$\text{F}_{EI}(\Delta_{EI}) = \int d^d x (a_{EI}|\Delta_{EI}|^2 + \frac{1}{2} b_{EI}|\Delta_{EI}|^4 + \text{K}_{EI}|\nabla\Delta_{EI}|^2),$$

(S2.3)

and $\text{F}_{SE}(\Delta_{SC}, \Delta_{EI}) = \int d^d x (c_{SE}|\Delta_{SC}|^2|\Delta_{EI}|^2),$

(S2.4)

with $a_{SC}$ and $a_{EI}$ being directly related to the susceptibility, $\text{K}_{SC}$ and $\text{K}_{EI}$ the rigidity of superconductor and exciton insulator, and $c_{SE}$ the coupling coefficient.

Near the critical point $T_c$, $\Delta_{SC}$ goes to zero, and $\Delta_{EI}$ can be considered as a constant, namely, $\Delta_{EI} = \Delta_0$. We then have

$$\mathrm{F}_{SC}(\Delta_{SC}) = \int d^d x (a_{SC}|\Delta_{SC}|^2 + \mathrm{K}_{SC}|\nabla\Delta_{SC}|^2)$$

$$= \sum_{\boldsymbol{k}} (a_{SC} + \mathrm{K}_{SC} k^2) |\Delta_{SC}(k)|^2, \qquad (S2.5)$$

where $\Delta_{SC}(x) = \frac{1}{\sqrt{S}} \sum_k \Delta_{SC}(k) e^{ikx}$ by performing the Fourier transformation of the superconducting gap. Similarly, the coupling term has the following expression,

$$\mathrm{F}_{SE}(\Delta_{SC}, \Delta_{EI}) =$$

$$\sum_{\boldsymbol{k_1 k_2 k_3 k_4}} c_{SE} \Delta^*_{SC}(\boldsymbol{k_1}) \Delta_{SC}(\boldsymbol{k_2}) \Delta^*_{EI}(\boldsymbol{k_3}) \Delta_{EI}(\boldsymbol{k_4}) \int e^{i(-\boldsymbol{k_1} + \boldsymbol{k_2} - \boldsymbol{k_3} + \boldsymbol{k_4}) \cdot \boldsymbol{x}} d^d \boldsymbol{x}$$

$$= \sum_{\boldsymbol{k_1 k_2 k_3 k_4}} c_{SE} \Delta^*_{SC}(\boldsymbol{k_1}) \Delta_{SC}(\boldsymbol{k_2}) \Delta^*_{EI}(\boldsymbol{k_3}) \Delta_{EI}(\boldsymbol{k_4}) \delta(-\boldsymbol{k_1} + \boldsymbol{k_2} - \boldsymbol{k_3} + \boldsymbol{k_4})$$

$$= \sum_{\boldsymbol{k}} c_{SE} |\Delta_{SC}(\boldsymbol{k})|^2 |\Delta_0|^2. \qquad (S2.6)$$

Finally, we obtain the effective free energy,

$$\mathrm{F}^{eff}(\Delta_{SC}) = \sum_{\boldsymbol{k}} (a_{SC} + \mathrm{K}_{SC} k^2 + c_{SE}|\Delta_0|^2) |\Delta_{SC}(\boldsymbol{k})|^2$$

$$= \sum_{\boldsymbol{k}} (\tilde{a}_{SC} + \mathrm{K}_{SC} k^2) |\Delta_{SC}(\boldsymbol{k})|^2, \qquad (S2.7)$$

where $\tilde{a}_{SC} = a_{SC} + c_{SE}|\Delta_0|^2$.

According to the discussions in Refs [S4,S5], the contribution due to quantum fluctuations is small when the following criterion is satisfied,

$$|\frac{T-T_c}{T_c}| \gg \frac{\alpha}{\mathrm{K}_{SC}}, \qquad (S2.8)$$

where $\alpha = \tilde{a}_{SC} \frac{T_c}{T-T_c}$.

To evaluate $\frac{\alpha}{\mathrm{K}_{SC}}$, we compute the parameter $\tilde{a}_{SC}$ by differentiating the Ginzburg-Landau free energy with respect to $\Delta_{SC}$, and obtain the parameter $\mathrm{K}_{SC}$ through the superfluid stiffness in this system. Using the parameters from the main text, we find that $\frac{\alpha}{K_{SC}}$ is typically very small near $T_c$, for example, $\frac{\alpha}{K_{SC}} \approx 0.08 * |\frac{T-T_c}{T_c}|$ at $T = 0.98T_c$, which ensures that the mean-field approximation can be adopted in our present study.

## 3. Total free energy

For distinct superconducting gap functions, the total energy of the system may vary and is easily influenced by the Coulomb attraction between electrons and holes. From Fig. S1, we can see that, for a given Coulomb attraction $V$, the energy for the *s*-wave pairing is the highest among several pairing states, and followed by slightly lower energies for the *d*±i*d*-wave and *f*-wave pairing phases. The discontinuous lines indicate that Cooper pairs are destroyed, leading to the potential disappearance of the corresponding superconducting phase. In contrast to the above pairing phases, the extended *s*-wave and *p*±i*p*-wave pairing states exhibit lower total energies and survive in the presence of strong Coulomb attraction. Besides, the energy difference between these two pairing phases decreases as $V$ increases.

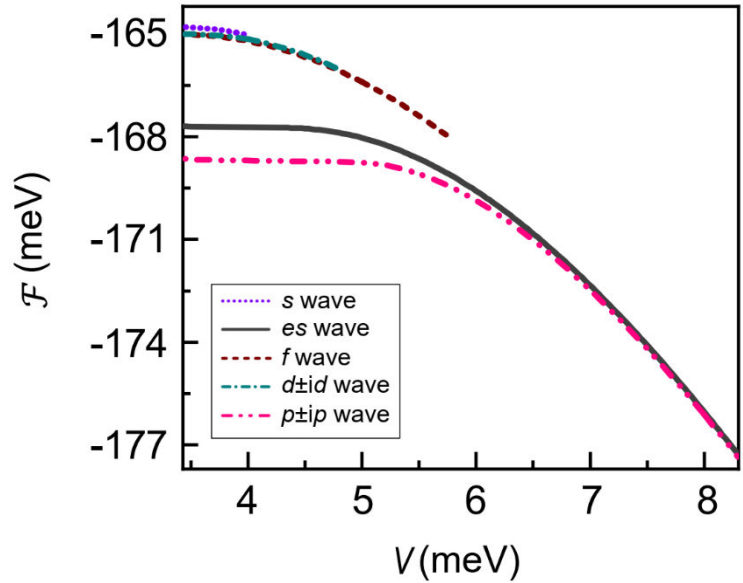

**Fig. S1:** Total energy of the system ($\mathcal{F}$) as a function of the Coulomb attraction ($V$) for the *s*-wave, extended *s*-wave (*es*-wave), *f*-wave, $d{\pm}id$-wave, and $p{\pm}ip$-wave pairing phases, respectively.

## Supplementary References